\documentclass[12pt]{article}
\pdfoutput = 1
\usepackage{amsmath,amssymb,bm,epsfig,afterpage}
\usepackage{mathtools}
\usepackage{cite}
\usepackage{here}
\usepackage{color}
\usepackage{tabularx}
\usepackage{simplewick} 
\usepackage{bbm}

\usepackage{comment}

\renewcommand{\thefootnote}{\fnsymbol{footnote}}

\newcommand{\abs}[1]{\left|{#1}\right|}

\newcommand{\eps}{\epsilon}

\newcommand{\hsg}{\hat{1}}
\newcommand{\hdb}{\hat{2}}
\newcommand{\htr}{\hat{3}}

\newcommand{\epth}[2]{\epsilon^{#1}\theta^{#2}} 
\newcommand{\order}[1]{\mathcal{O}\left({#1}\right)}

\newcommand{\pr}{\prime}

\newcommand{\SL}[2]{\mathrm{SL}({#1},{#2})}

\newcommand{\intZ}{\mathbb{Z}}
\newcommand{\oGam}{\overline{\Gamma}} 
 
\newcommand{\Ita}{\mathrm{Im}\,\tau}

\newcommand{\CKM}{\mathrm{CKM}}

\def\Im{\mathop{\mathrm{Im}}}

\allowdisplaybreaks[3]
\newcolumntype{Y}{&gt;{\centering\arraybackslash}X} 
\usepackage{hyperref}

\begin{document}

\begin{titlepage}

\begin{flushright}
 {\tt
CTPU-PTC-23-01 \\
EPHOU-23-002 
}
\end{flushright}

\vspace{1.2cm}
\begin{center}
{\Large
{\bf
Quark masses and CKM hierarchies \\ from $S_4'$ modular flavor symmetry  
}
}
\vskip 2cm
Yoshihiko Abe$^a$~\footnote{yabe3@wisc.edu}, 
Tetsutaro Higaki$^b$~\footnote{thigaki@rk.phys.keio.ac.jp}, 
Junichiro Kawamura$^{b,c}$~\footnote{junkmura13@gmail.com}
and
Tatsuo Kobayashi$^d$~\footnote{kobayashi@particle.sci.hokudai.ac.jp}

\vskip 0.5cm

{\it $^a$
Department of Physics, University of Wisconsin, Madison, WI 53706, USA
}\\[3pt]

{\it $^b$
Department of Physics, Keio University, Yokohama, 223-8522, Japan
}\\[3pt]

{\it $^c$
Center for Theoretical Physics of the Universe, Institute for Basic Science (IBS),
Daejeon 34051, Korea
}\\[3pt]

{\it $^d$
Department of Physics, Hokkaido University, Sapporo 060-0810, Japan}\\[3pt]

\vskip 1.5cm

\begin{abstract}
We propose models to explain the hierarchies of the quark masses and mixing 
by utilizing the $S_4^\prime$ modular flavor symmetry. 
The hierarchy is realized by the modulus $\tau$ stabilized at $\mathrm{Im}\,\tau \gg 1$, 
where the residual $\mathbbm{Z}_4^T$ symmetry is approximately unbroken 
and the Froggatt-Nielsen mechanism works. 
It is found that the quark hierarchies are realized only in a few cases of quark representations. 
We study two models with assigning the modular weights, 
so that the observed quark hierarchies are explained in the cases of  
both small and large ratios of the top to bottom Yukawa couplings. 
We also argue that $\order{0.1}$ hierarchies of the $\order{1}$ coefficients  
and the spontaneous CP violation 
can be realized by imposing another $S_3$ modular symmetry. 
\end{abstract}
\end{center}
\end{titlepage}

%\tableofcontents
\clearpage

\renewcommand{\thefootnote}{\arabic{footnote}}
\setcounter{footnote}{0}

\section{Introduction}

Understanding the origin of the flavor structure of quarks and leptons is one 
of the big challenges in particle physics.
Recently, the modular flavor symmetry attracts the attention as an interesting possibility to explain the flavor structure~\cite{Feruglio:2017spp}.
In these models, 
the three generations of quarks and leptons transform non-trivially under the modular symmetry, 
that is, the modular symmetry is in a sense a flavor symmetry.
On top of that, Yukawa couplings are assumed to be modular forms, 
which are holomorphic functions of the modulus $\tau$ 
and non-trivially transform under the action of the modular group.
As discussed in Ref.~\cite{deAdelhartToorop:2011re}, it is remarkable that the (in)homogeneous finite modular group $\Gamma_N^{(\prime)}$ with the level $N \leq 5$ is isomorphic to the well-known (double-covering of) permutation group, 
such as $S_3$, $A_4^{(\prime)}$, $S_4^{(\prime)}$ and $A_5^{(\prime)}$, which have been intensively studied to explain the lepton flavor structure in the literature~\cite{Feruglio:2017spp,Kobayashi:2018vbk,Penedo:2018nmg,Novichkov:2018nkm,Ding:2019xna,Liu:2019khw,Novichkov:2020eep,Liu:2020akv,Liu:2020msy}.
These non-Abelian finite groups have been studied in flavor models for quarks and leptons~\cite{Altarelli:2010gt,Ishimori:2010au,Ishimori:2012zz,Hernandez:2012ra,King:2013eh,King:2014nza,Tanimoto:2015nfa,King:2017guk,Petcov:2017ggy,Feruglio:2019ybq,Kobayashi:2022moq}.
The phenomenological aspects of the modular flavor symmetries 
have been actively discussed in the literature~\cite{Criado:2018thu,Kobayashi:2018scp,Ding:2019zxk,Novichkov:2018ovf,Kobayashi:2019mna,Wang:2019ovr,Chen:2020udk,deMedeirosVarzielas:2019cyj,Asaka:2019vev,Asaka:2020tmo,deAnda:2018ecu,Kobayashi:2019rzp,Novichkov:2018yse,Kobayashi:2018wkl,Okada:2018yrn,Okada:2019uoy,Nomura:2019jxj,Okada:2019xqk,Nomura:2019yft,Nomura:2019lnr,Criado:2019tzk,King:2019vhv,Ding:2019gof,deMedeirosVarzielas:2020kji,Zhang:2019ngf,Nomura:2019xsb,Kobayashi:2019gtp,Lu:2019vgm,Wang:2019xbo,King:2020qaj,Abbas:2020qzc,Okada:2020oxh,Okada:2020dmb,Ding:2020yen,Okada:2020rjb,Okada:2020ukr,Nagao:2020azf,Wang:2020lxk,Okada:2020brs,Yao:2020qyy,Kuranaga:2021ujd}.

The modular symmetry is well-motivated from the higher dimensional theories such as superstring theory.
For example, if we consider the torus or its orbifold compactification, the modulus parameter $\tau$ is the complex structure modulus, which is a dynamical degree of freedom of the effective field theory determining the shape of the torus.
The modular symmetry appears as the geometrical symmetry associated with this compact space.
The Yukawa couplings are obtained by the overlap integral of the profile functions of the matter zero-modes and expressed as the function of the modulus which transform non-trivially under the modular transformation.
The behavior of the zero-mode function under the modular transformation was studied in magnetized D-brane models~\cite{Kobayashi:2017dyu,Kobayashi:2018rad,Kobayashi:2018bff,Ohki:2020bpo,Kikuchi:2020frp,Kikuchi:2020nxn,Hoshiya:2020hki} and heterotic orbifold modelds~\cite{Lauer:1989ax,Lauer:1990tm,Ferrara:1989qb,Baur:2019kwi,Nilles:2020nnc,Nilles:2020gvu}.
The modular flavor symmetric three-generation models based on the magnetized extra dimension were discussed in Refs.~\cite{Hoshiya:2020hki,Kikuchi:2021ogn}.
The modulus stabilization is also discussed in Refs.~\cite{Ishiguro:2020tmo,Novichkov:2022wvg}.

A certain residual symmetry remains unbroken 
when the modular symmetry is broken by the vacuum expectation value (VEV) 
of the modulus at a certain fixed point. 
The residual symmetry $\mathbb{Z}^{S}_{4} \subset \Gamma_N^\pr$ 
or $\mathbb{Z}^S_{2} \subset \Gamma_N$, associated with the $S$ generator, 
remains unbroken at $\tau=i$.   
In addition, $\mathbb{Z}^{ST}_3$ and $\mathbb{Z}_N^{T}$, associated with $ST$ and $T$,  
remains at $\tau = \omega := e^{2\pi i/3}$ and $\tau = i\infty$, respectively. 
These residual symmetries have been utilized in model-building in the literature~\cite{Novichkov:2018ovf,Novichkov:2018yse,Novichkov:2018nkm,Okada:2020brs}.
It is particularly interesting that the hierarchical structure 
of Yukawa matrices for the Standard Model (SM) fermions 
can be realized at a vicinity of the fixed points.  
Indeed the lepton sector was discussed in Refs.~\cite{Feruglio:2021dte,Novichkov:2021evw}.

In this paper, we discuss the modular flavor symmetry which can realize the hierarchical structures of the quark masses and the Cabbibo-Kobayashi-Maskawa (CKM) mixing.
The realizations of the quark mass hierarchy were discussed by use of $\Gamma_3 \simeq A_4$ at $\tau \sim \omega$ and $\Gamma_6$ at $\Ita \gg 1$  in Refs.~\cite{Petcov:2022fjf} and \cite{Kikuchi:2023cap}, respectively.
We focus on the modular flavor symmetry at $N=4$, which is isomorphic to $S_4^{(\prime)}$.
This is the minimal possibility to realize the hierarchical structure 
with up to cubic order of a small parameter which may be necessary to 
explain the quark hierarchies.
For $\Im \tau \gg 1$, this model has approximately the residual discrete symmetry $\mathbb{Z}^T_4$, which realizes the quark mass structures 
by the Froggatt-Nielsen (FN) mechanism~\cite{Froggatt:1978nt,Higaki:2019ojq}.
It will be turned out that a few patterns of representations 
can realize the quark hierarchical structure. 
We then explicitly construct two models with assigning modular weights, 
so that the experimental values of the quark masses and the CKM angles 
are explained with small and $\order{1}$ ratios of bottom to top quark Yukawa couplings.

The rest of the paper is organized as follows.
In Sec.~\ref{sec-N=4-modular-form}, we briefly review the modular symmetry at $N=4$.
We discuss the textures of the modular forms at $\Ita \gg 1$, where the the residual $\mathbb{Z}^T_4$ symmetry is realized. 
In Sec.~\ref{sec-hierarchical-structures}, 
we discuss possible quark representations to realize  the quark mass hierarchy.
We study two models with different modular weights based on the modular flavor symmetry $S_4^{(\prime)}$ in Sec.~\ref{sec-model}. 
Sec.~\ref{sec-summary} is devoted to summary. 
The details of $S_4^\pr$ modular flavor symmetry as well as $S_3$ are shown in App.~\ref{app-S4p}.

\section{Modular symmetry at $N=4$} 
\label{sec-N=4-modular-form}

We briefly review the modular symmetry.
The homogeneous modular group $\Gamma \coloneqq \SL{2}{\intZ}$ is defined as 
\begin{align}
\Gamma := \left\{
    \begin{pmatrix}
            a & b \\
            c & d
    \end{pmatrix}
 \ 
 \Biggl|
 \ 
 a,b,c,d \in \intZ, 
 \quad 
 ad-bc = 1 
        \right\}. 
\end{align}
This group is generated by the generators 
\begin{align}
 S=\begin{pmatrix}
            0 & 1 \\
            -1 & 0
        \end{pmatrix} , \qquad 
 T=\begin{pmatrix}
            1 & 1 \\
            0 & 1
        \end{pmatrix} , \qquad
 R=\begin{pmatrix}
            -1 & 0 \\
            0 & -1
        \end{pmatrix} ,
\end{align}
and they satisfy the following algebraic relations,
\begin{align}
    S^2 = R,
    \qquad 
    (ST)^3 = R^2 = \mathbb{I},
    \qquad 
    TR = RT.
\end{align}
The inhomogeneous modular group $\oGam \coloneqq {\rm P}\SL{2}{\intZ}$ is 
defined by $\oGam :=\Gamma/\mathbb{Z}^R_2$, where $\mathbb{Z}^R_2$ is generated by $R$.
That is, the generator $R$ is presented by $R=\mathbb{I}$ in  $\oGam := {\rm P}\SL{2}{\intZ}$.

In addition, congruence subgroup $\Gamma(N)$ is  defined by 
\begin{align}
    \Gamma(N) \coloneqq \left\{
        \begin{pmatrix}
            a & b \\
            c & d
        \end{pmatrix} \in \SL{2}{\intZ},
        \quad 
        \begin{pmatrix}
            a & b \\
            c & d
        \end{pmatrix} \equiv 
        \begin{pmatrix}
            1 & 0 \\
            0 & 1
        \end{pmatrix}
        \mod N
    \right\}.
\end{align}
The quotients $\Gamma_N:= \oGam/\Gamma(N)$
for $N=2,3,4$ and $5$ are respectively isomorphic to $S_3, A_4, S_4$ and $A_5$.
Moreover, the quotients $\Gamma^\pr_N:=\Gamma/\Gamma(N)$ 
for $N=3,4$ and $5$ are isomorphic to $A_4^\pr, S_4^\pr$ and $A_5^\pr$, 
which are double covering groups of $A_4, S_4$ and $A_5$, respectively.
In these quotients, the generator $T$ satisfies 
\begin{align}
    T^N=\mathbb{I}, 
\end{align}
and thus it generates $\mathbbm{Z}_N^T$ symmetry.

The group $\Gamma_4' \simeq S_4'$ has 10 irreducible representations,
\begin{align}
 1,1^\pr, 2, 3, 3^\pr
\quad\text{and}\quad
 \hsg, \hsg^\pr, \hdb, \htr, \htr^\pr. 
\end{align}
The non-hatted representations $r$ are those in the $S_4$ symmetry, 
transformed by $R$ trivially, i.e. $R\,r = r$, 
while the hatted representations $\hat{r}$ are transformed non-trivially by $R$, 
i.e. $R\,\hat{r}= -\hat{r}$.  
Throughout this work, we use the representation matrices 
in which matrices are diagonal for $T$ and real for $S$, 
shown in App.~\ref{app-S4p}.

The modular group $\Gamma$ acts on the modulus $\tau$ ($\Ita >0$)
as 
\begin{align}
 \tau \to \frac{a\tau + b}{c\tau+d}. 
\end{align} 
A modular form $Y_r^{(k)}$ of representation $r$ under $\Gamma_4^\pr$  
with a weight $k$ transforms as 
\begin{align}
 Y_{r}^{(k)}(\tau) \to (c\tau+d)^{k} \rho(r) Y_r^{(k)}(\tau),    
\end{align}
where $\rho(r)$ is the representation matrix. 
The number of representations at a weight $k$ is listed in Table~\ref{tab-wgt} of App.~\ref{app-modform}.
At $k=1$, there is a $\htr$ representation,  
\begin{align}
 Y^{(1)}_{\htr}(\tau) = 
\begin{pmatrix}
 \sqrt{2}\eps(\tau) \theta(\tau) \\ \eps^2(\tau) \\ -\theta^2(\tau)
\end{pmatrix},  
\end{align}
where the functions $\theta$ and $\eps$ are written 
by the Jacobi theta functions~\cite{Novichkov:2020eep}. 
Their series forms are given by 
\begin{align}
 \theta(\tau) = 1+2\sum_{n=1}^\infty q^{n^2}, 
\quad 
 \eps(\tau) = 2q^{1/4} \sum_{n=0}^\infty q^{n(n+1)},  
\end{align}  
where $q := e^{2\pi i\tau}$. 
The modular forms at higher weights can be constructed from products of $Y^{(1)}_{\htr}$, 
and the ones used in our models are shown in App.~\ref{app-modform}.

At $\Ita\gg 1$, $\theta \sim 1$ and $\abs{\eps} \sim 2 e^{-(\pi/2) \Ita} \ll 1$, 
and hence $\eps$ will be the origin for the quark hierarchies.
In this limit, the symmetry generated by $T$ is unbroken, 
and thus the $\mathbbm{Z}^T_4$ symmetry can realize the FN-like mechanism~\cite{Froggatt:1978nt,Higaki:2019ojq}, 
where the flavon is replaced by $\eps(\tau) \sim 2 q^{1/4}$ 
whose $\mathbbm{Z}^T_4$ charge is one. 
The irreducible representations have the following hierarchical structures in this limit,
\begin{align}
\label{eq-Yhie}
& Y_1 \sim 1, \quad Y_{1^\pr} \sim  \eps^2, \quad 
Y_{\hsg} \sim \eps^3, \quad Y_{\hsg^\pr} \sim \eps, 
\quad Y_{2} \sim  
\begin{pmatrix}
 1 \\ \eps^2 
\end{pmatrix}, 
\quad 
Y_{\hdb} \sim 
\begin{pmatrix}
 \eps^3 \\ \eps  
\end{pmatrix}, 
\\ \notag 
& 
Y_{3} \sim 
\begin{pmatrix}
 \eps^2 \\ \eps^3 \\ \eps  
\end{pmatrix}, 
\quad 
Y_{3^\pr} \sim 
\begin{pmatrix}
 1 \\ \eps \\ \eps^3 
\end{pmatrix}, 
\quad 
Y_{\htr} \sim 
\begin{pmatrix}
 \eps \\ \eps^2 \\ 1 
\end{pmatrix}, 
\quad 
Y_{\htr^\pr} \sim 
\begin{pmatrix}
 \eps^3 \\ 1 \\ \eps^2 
\end{pmatrix}, 
\end{align}
where $Y_r$ is the modular form of the representation $r$. 
The weights $k$ are omitted here 
since the hierarchical structures are determined only by the $\mathbbm{Z}_4^T$ charge 
and are independent of the weight for a given representation (see the representation matrix $\rho(r)$ shown in App.~\ref{app-modform}). 
We see that the maximum power of $\eps$ is $N-1 = 3$ 
which may be the minimal number to explain the quark hierarchies \cite{Higaki:2019ojq}.

\section{Hierarchical structures}  
\label{sec-hierarchical-structures}
 
The goal of this work is to explain the hierarchical structure 
in the quark sector with ${\cal O}(1)$ free parameters.  
The quark hierarchies may be expressed by a small parameter $\eps\ll 1$, 
\begin{align}
\label{eq-texture} 
 (m_u, m_c, m_t) \sim (\eps^3, \eps, 1), 
\quad 
 (m_d, m_s, m_b) \sim 
\eps^p\times(\eps^2, \eps^2, 1), 
\quad 
V_\CKM 
\sim 
\begin{pmatrix}
 1 & 1 & \eps^{2} \\
 1 & 1 & \eps^{2} \\
 \eps^{2} &\eps^2 & 1 
\end{pmatrix},   
\end{align}
where $p=0,1$. 
The top to bottom mass ratio $m_t/m_b$ will be explained 
by $\eps$ for $p=1$, while, for $p=0$, it is explained by $\tan\beta := v_u/v_d$, 
where $v_u$ ($v_d$) being the VEV 
of the neutral component of the up-type (down-type) Higgs doublet in 
two Higgs doublet models, such as supersymmetric models.
We note that $N=4$ is the minimum possibility to realize the texture 
in Eq.~\eqref{eq-texture}, since the maximum power of $\eps$ is $N-1$: $\epsilon^{N-1} = \epsilon^{3} \sim m_u/m_t$. 
One may think that the texture may not fully fit the data, 
especially for the strange to down quark mass ratio $m_s/m_d$
and the CKM angles involving the third generation. 
The former is predicted to be $\order{1}$ and the latter may be too small. 
It is shown in Ref.~\cite{Higaki:2019ojq}
 that the CKM angles with $\order{\eps}$ fits to the data.  
We will see later that these potential issues are resolved 
by the canonical normalizations and the numerical coefficients in the modular forms. 
Recently, the quark hierarchical structures realized 
by the level $N=3$ at $\tau\simeq \omega$ and $N=6$ at $\tau \simeq i\infty$ 
are studied in Refs.~\cite{Petcov:2022fjf} and~\cite{Kikuchi:2023cap}, 
respectively.

The hierarchical structure of the masses from the modular flavor symmetry 
is listed in Ref.~\cite{Novichkov:2021evw}~\footnote{
In this work, we consider the combinations of the representations shown in Ref.~\cite{Novichkov:2021evw}, 
i.e. the hatted and non-hatted representations do not appear in the same type of quarks. 
}.
There are several patterns which realize the up and down quark masses, 
but not all of them realize the CKM matrix. 
For instance, if the $SU(2)_L$ doublet quark $Q$ is a $S_4'$ triplet of any kind, 
the up-type Yukawa couplings are originated 
from $Y_{3^\pr}$, see Eq.~\eqref{eq-Yhie}, 
so that the largest element is $\order{1}$ 
and the up quark mass hierarchy is realized. 
The down-type Yukawa couplings are originated from 
$Y_3$ ($Y_{\htr}$) if $p=1$ ($p=0$), 
so that the smallest and largest elements differ by $\eps^2$. 
In this case, 
the top quark is predominantly from the first element, 
while the bottom quark is predominantly from the second or third element,  
and thus $\abs{V_{tb}} \ll 1$ is predicted. 
In addition, the Cabbibo angle $\sim \abs{V_{us}}$ 
is expected to be $\order{\eps}$ 
because of the hierarchical structure in the triplets.
If $Q$ is composed of the same singlet, e.g. $Q = 1\oplus 1\oplus 1$, 
all of the mixing angles are predicted to be $\order{1}$, 
and thus the hierarchical pattern is not explained~\footnote{
The down quark mass hierarchy $(\eps^3, \eps^2, \eps)$ is incompatible with the CKM hierarchy 
for the same reason, 
%The doublet quark $Q$ should not be triplets or composed of the same singlet, 
because the mass hierarchy is realized only by $3\otimes (1\oplus 1\oplus 1)$. 
%because it is realized by the combination of $3\otimes (1\oplus 1\oplus 1)$, 
%and thus $Q$ should be neither of them which predicts $\abs{V_{tb}}\ll 1$ or $\order{1}$ mixing.  
%Thus, the down quark mass hierarchy then given by $(\eps^3, \eps^2, \eps)$ is not compatible with the CKM matrix. 
}.  
Therefore, the texture in Eq.~\eqref{eq-texture} is realized 
only in the following four cases:   
\begin{align}
\label{eq-creps}
u^c = 3, \quad 
d^c = 
\begin{cases}
1^\pr \oplus 1^\pr \oplus 1^\pr 
\\
\hsg^\pr\oplus\hsg^\pr\oplus\hsg^\pr  
\end{cases}, 
\quad 
Q =
\begin{cases}
  2\oplus 1 \\  1^\pr\oplus 1\oplus 1 
\end{cases}.  
\end{align}
Note that the cases such as  
$u^c = 3^\pr$, $d^c = 1\oplus 1\oplus1$ and $Q = 2^\pr \oplus 1^\pr$ 
give the same Yukawa structure, 
so the phenomenology will not be changed from the above four cases.  
The first (second) case for $d^c$ corresponds to $p=0$ ($p=1$). 
The texture is the same for the two cases of $Q$.  
We shall study the first case since it is more predictive, 
because of the smaller number of parameters. 
The second case is obtained 
by splitting $2$ into $1^\pr \oplus 1$ in the first case.

\section{Models}
\label{sec-model}

\begin{table}[t]
 \centering
\caption{\label{tab-mat} 
Assignments of the quarks and Higgs doublets 
under $G_{\mathrm{EW}} := SU(2)_L\times U(1)_Y$, $S_4^\pr$ and the weight $k$. 
}
\begin{tabular}[t]{c|c|ccc|cc|cc} \hline 
     & $u^c$ & $d^c_1$ & $d^c_2 $ & $d^c_3$ & $q_1$ & $q_2$ & $H_u$ & $H_d$ 
\\ \hline \hline 
$G_{\mathrm{EW}}$ & $1_{-2/3}$ & \multicolumn{3}{c|}{$1_{1/3}$}  
                      & \multicolumn{2}{c|}{$2_{1/6}$} & $2_{1/2}$ & $2_{-1/2}$  \\ \hline
$S_4^\pr$ & $3$ & \multicolumn{3}{c|}{$\hsg^\pr$ or $1^\pr$} & $1$ & $2$ & $1$ & $1$\\  
$k$        & $-k_u$ & $-k_{d_1}$ & $-k_{d_2}$ & $-k_{d_3}$ & $-k_{q_1}$ & $-k_{q_2}$ & $0$ & 0 \\ \hline  
\end{tabular}
\end{table}

 We construct supersymmetric models with the representations shown in Eq.~\eqref{eq-creps} 
 which realizes the texture in Eq.~\eqref{eq-texture}. 
The assignments of the chiral superfields under the electroweak (EW) gauge symmetry, $S_4^\pr$ 
and the modular weights $k$ are shown in Table~\ref{tab-mat}. 
For general weight assignments, the Yukawa couplings are given by
\begin{align}
 W =&\ H_u \left\{ \alpha_1 q_1 \left(Y_3^{(k_u + k_{q_1})} u^c\right)_1 
         + \alpha_2 \left(q_2 Y^{(k_u + k_{q_2})}_3 u^c \right)_1 
            +  \alpha_3 \left(q_2 Y^{(k_u+k_{q_2})}_{3^\pr} u^c \right)_1\right\}  \\ \notag 
        & \quad 
          + H_d \sum_{i=1}^3 
 \left\{ \beta_{1i} q_1 \left(Y^{(k_{d_i}+k_{q_1})}_{\mathbf{1}} d^c_i \right)_1 
        +\beta_{2i} \left( q_2 Y^{(k_{d_i}+k_{q_2})}_{\mathbf{2}}d^c_i\right)_{1}      
\right\}\\ \notag 
=:&\ H_u Q Y_u u^c + H_d Q Y_d d^c,         
\end{align}
where $(\cdots)_{1}$ 
is the trivial singlet combination of the product inside the parenthesis~\footnote{
The products of the irreducible representations of $S_4^\pr$ 
are listed in App.~\ref{app-S4p}. 
}.  
Besides, there will be more coefficients 
if there are more than one modular forms which are degenerate 
for a given representation $r$ and weight $k$. 
On the contrary, the term is understood to be absent 
if there is no modular form for a given $r$ and $k$. 
For explicit examples, see the models in the following sections.  
For the down quark couplings, $(\mathbf{1}, \mathbf{2}) = (1^\pr, 2)$ 
and $(\hsg, \hdb)$ for $p=0$ and $p=1$, respectively.  
In the second line, we defined 
$Q := (Q_1, Q_2, Q_3)$, $u^c := (u^c_1, u^c_2, u^c_3)$ 
and $d^c := (d^c_1, d^c_2, d^c_3)$. 
We assign $Q_1$ is the singlet and the others forming the doublet under $S_4^\pr$, 
i.e. $q_1 := Q_1$, $q_2 := (Q_2, Q_3)$. 
The up-type quark $u^c$ is the triplet and each element of $d^c$ is the singlet.  
The K\"{a}hler potential of the quark chiral superfield $q$ with wight $k_q$, 
which includes the kinetic term, is given by
\begin{align}
K\supset \frac{q^\dagger q}{(-i\tau+i\tau)^{k_q}},   
\end{align}
then, after the canonical normalization, the Yukawa matrices are normalized as 
\begin{align}
\label{eq-cnorm}
 \left[Y_u\right]_{ij} \to\left(\sqrt{2\,\Ita}\right)^{k_{q_i} + k_u} \left[Y_u\right]_{ij}, 
\quad 
 \left[Y_d\right]_{ij} \to\left(\sqrt{2\,\Ita}\right)^{k_{q_i} + k_{d_j}} \left[Y_d\right]_{ij}, 
\end{align}
where $i,j = 1,2,3$ and $k_{q_3} = k_{q_2}$.
At $\Ita \gg 1$ where $\eps \sim \order{0.01}$,  
this normalization factor $2\,\Ita \sim 5$ can be important for the hierarchical structure.

The hierarchical structure of the Yukawa matrices before the canonical normalization 
are given by 
\begin{align}
 Y_u P_{13} \sim 
\begin{pmatrix}
\eps^3 & \eps &  \eps^2 \\
\eps^3 & \eps &  \eps^2 \\
\eps & \eps^3 & 1 
\end{pmatrix}, 
\quad 
 Y_d P_{13} \sim \eps^p 
\begin{pmatrix}
\eps^2 & \eps^2 & \eps^2 \\ 
\eps^2 & \eps^2 & \eps^2 \\ 
1   & 1   &  1 \\ 
\end{pmatrix}, 
\quad 
\mathrm{where}
\quad 
P_{13}:= 
\begin{pmatrix}
 0 & 0 & 1 \\
 0 & 1 & 0 \\
 1 & 0 & 0 \\
\end{pmatrix}. 
\end{align}
Here, $P_{13}$ is multiplied so that $(3,3)$ element 
is predominantly the top and bottom Yukawa couplings. 
These structures realize the mass and CKM hierarchies in Eq.~\eqref{eq-texture}.    
We shall consider the two models 
which can explain the quark hierarchies for $p=0$ and $p=1$.

\subsection{Large $\tan\beta$ scenario: $p=0$}  
\label{sec-mdlA}

\begin{table} 
\centering 
\caption{\label{tab-obs}
The values of the Yukawa couplings at benchmark points 
in the case of $p=0$ (left) and $p=1$ (right). 
The second column is predictions of our models, 
and the third (fourth) column shows the experimental values (its $1\sigma$ error).
The central values are at the GUT scale 
after the renormalization group evolution from the experimental values 
when $M_{\mathrm{SUSY}} = 10~\mathrm{TeV}$ 
and vanishing threshold corrections in the MSSM~\cite{Antusch:2013jca}.  
The errors at the scale $10$ TeV are shown for reference. 
} 
\begin{minipage}[t]{0.48\textwidth}
\centering 
\begin{tabular}[t]{c|ccc} \hline 
obs. & value & center & error \\  \hline \hline 
$y_u$$/10^6$ & 2.8 & 2.7 & 1.3 \\  
$y_c$$/10^3$ & 1.487 & 1.422 & 0.095 \\  
$y_t$ & 0.5139 & 0.5139 & 0.0084 \\  \hline 
$y_d$$/10^4$ & 1.9935 & 1.9935 & 0.0087 \\  
$y_s$$/10^3$ & 3.946 & 3.946 & 0.014 \\  
$y_b$ & 0.2282 & 0.2282 & 0.0001 \\  \hline 
$s_{12}$ & 0.2274 & 0.2274 & 0.0007 \\  
$s_{23}$$/10^2$ & 3.945 & 3.942 & 0.065 \\  
$s_{13}$$/10^3$ & 3.43 & 3.43 & 0.13 \\  
$\delta_{\mathrm{CP}}$ & 1.215 & 1.208 & 0.054 \\  \hline 
\end{tabular} 
\end{minipage}
\begin{minipage}[t]{0.48\textwidth}
\centering 
\begin{tabular}[t]{c|ccc} \hline 
obs. & value & center & error \\  \hline \hline 
$y_u$$/10^6$ & 2.9 & 2.9 & 1.3 \\  
$y_c$$/10^3$ & 1.560 & 1.508 & 0.095 \\  
$y_t$ & 0.5464 & 0.5464 & 0.0084 \\  \hline 
$y_d$$/10^6$ & 9.00 & 9.06 & 0.87 \\  
$y_s$$/10^4$ & 1.73 & 1.79 & 0.14 \\  
$y_b$$/10^2$ & 1.011 & 0.994 & 0.013 \\  \hline 
$s_{12}$ & 0.2274 & 0.2274 & 0.0007 \\  
$s_{23}$$/10^2$ & 3.991 & 3.989 & 0.065 \\  
$s_{13}$$/10^3$ & 3.47 & 3.47 & 0.13 \\  
$\delta_{\mathrm{CP}}$ & 1.204 & 1.208 & 0.054 \\  \hline 
\end{tabular} 
\end{minipage}
\end{table}

We assign the modular weights as 
\begin{align}
k_{q_1} = 2, \quad k_{q_2} = 4, \quad k_u = 2, 
\quad 
k_{d_1} = 4, \quad k_{d_2} = 2, \quad k_{d_3} = 0. 
\end{align}
Since there is no odd weight, there are only non-hatted representations. 
This means that the inhomogeneous group $\Gamma_4^\pr/R \simeq S_4$ is enough for this model.
We can also study other patterns of the weights where the weights of the Yukawa couplings 
are less than 10, but this setup has the smallest hierarchy among the parameters 
with explaining the experimental values.

In this case, the superpotential is 
\begin{align}
W =&\ H_u \left\{ \alpha_1 q_1 \left(Y_3^{(4)} u^c\right)_1 
         + \alpha_2 \left(q_2 Y^{(6)}_3 u^c \right)_1 
            + \sum_{i_Y=1}^2 \alpha_3^{i_Y} \left(q_2 Y^{i_Y(6)}_{3^\pr} u^c \right)_1 
            \right\}  \\ \notag 
        & \quad 
          + H_d  
 \left\{ \beta_{11} q_1 \left(Y^{(6)}_{{1}^\pr} d^c_1 \right)_1 
           + \sum_{i_Y=1}^2 \beta_{21}^{i_Y} \left( q_2 
        Y^{i_Y(8)}_{{2}}d^c_1\right)_{1}  
        + \sum_{j=2}^3 \beta_{2j} \left( q_2 Y^{(10-2j)}_{2}d^c_j\right)_{1} 
\right\}.     
\end{align}
The Yukawa matrices are given by 
\begin{align}
 Y_u = &\ 
\begin{pmatrix}
 \alpha_1 [Y^{(4)}_3]_1 & 
 \alpha_1 [Y^{(4)}_3]_3 & 
 \alpha_1 [Y^{(4)}_3]_2 \\ 
-2 \alpha_2 [Y^{(6)}_3]_1 & 
   \alpha_2 [Y^{(6)}_3]_3 +\sqrt{3}\alpha_3^{i_Y} [Y^{i_Y(6)}_{3^\pr}]_2 & 
   \alpha_2 [Y^{(6)}_3]_2 +\sqrt{3}\alpha_3^{i_Y} [Y^{i_Y(6)}_{3^\pr}]_3 \\ 
-2 \alpha_3^{i_Y} [Y^{i_Y(6)}_{3^\pr}]_1 & 
   \alpha_3^{i_Y} [Y^{i_Y(6)}_{3^\pr}]_3 -\sqrt{3}\alpha_2 [Y^{(6)}_{3}]_2 & 
   \alpha_3^{i_Y} [Y^{i_Y(6)}_{3^\pr}]_2 -\sqrt{3}\alpha_2 [Y^{(6)}_{3}]_3 \\ 
\end{pmatrix}, 
\notag \\
Y_d =&\
\begin{pmatrix}
\beta_{11} Y^{(6)}_{1^\pr}     & 0 & 0 \\ 
-\beta_{21}^{i_Y} [Y^{i_Y(8)}_{2}]_2 
&-\beta_{22} [Y^{(6)}_{2}]_2&-\beta_{23}[Y^{(4)}_{2}]_2  \\ 
\beta_{21}^{i_Y} [Y^{i_Y (8)}_{2}]_1 
& \beta_{22} [Y^{(6)}_{2}]_1& \beta_{23}[Y^{(4)}_{2}]_1 
\end{pmatrix}, 
\end{align}
where $[Y^{(k)}_r]_{i}$ is the $i$-th element of $Y^{(k)}_r$. 
Here, the summation over $i_Y = 1,2$ is implicit. 
Note that there is no $1^\pr$ at $k=2,4$, so $(1,2)$ and $(1,3)$ elements in $Y_d$ are zero.
Altogether there are 9 coefficients, 
namely $\alpha_1,\alpha_2,\alpha_3^1, \alpha_3^2,\beta_{11},\beta_{12}^1, \beta_{12}^2, \beta_{22}$ 
and $\beta_{23}$. 
We introduce a phase to $\alpha_3$,  
since the CKM phase is approximately vanishing if all of the coefficients are real.

We fit $\tan\beta$, $\tau$ and the 9 coefficients to explain 
the observed quark masses and the CKM mixing. 
The values of Yukawa coupling constants and CKM angles at the Grand Unified Theory (GUT)
scale in the MSSM are calculated in Ref.~\cite{Antusch:2013jca}. 
Throughout our analysis, we refer to the values when $M_{\mathrm{SUSY}} = 10$~TeV, 
and the SUSY threshold corrections are zero~\footnote{
The values in $1 < \tan\beta < 5$ are linearly extrapolated from $\tan\beta > 5$.   
}.  
We find the benchmark point where the observables are within $1\sigma$ range,
as shown in the left panel of Table~\ref{tab-obs}. 
The input parameters at this point are  
$\tan\beta = 36.0982$, $\tau = 2.4956+2.2306i$, 
$|\alpha_3^1| = 4.1369\times 10^{-3}$ and 
\begin{align}
\frac{1}{|\alpha_3^1|} 
\begin{pmatrix} 
\alpha_{1} \\ \alpha_{2} \\ \alpha_{3}^1 \\ \alpha_3^2 
\end{pmatrix} 
= 
\begin{pmatrix} 
-0.1357 \\ -1.6734 \\ e^{0.0074i} \\ -0.6894 
\end{pmatrix} 
, \quad 
\frac{1}{|\alpha_3^1|} 
\begin{pmatrix} 
\beta_{11} \\ \beta_{21}^1 \\ \beta_{21}^2 \\ \beta_{22} \\ \beta_{23} 
\end{pmatrix} 
= 
\begin{pmatrix} 
-3.1165 \\ 0.1350 \\ 1.6214 \\ -0.1357 \\ 0.2806 
\end{pmatrix}. 
\end{align}
Here, we normalize the coefficients by the absolute value of $\alpha_3^1$. 
The overall factor for the coefficients are needed to be small 
to compensate the relatively large factor from the canonical normalization in Eq.~\eqref{eq-cnorm}. 
Although this would not be a problem 
because we do not know the overall normalization of the Yukawa forms,
the smallness of overall factor of the coefficients could also
be explained by other moduli, which appear only in the overall factor~\cite{Cremades:2004wa}. 
The ratio of the largest and smallest absolute values of the coefficients is about $23$.

With the assignment of the weights, 
the predictions for the quark masses and CKM angles at $t := 2\,\Ita \gg 1$ are modified as 
\begin{align}
\label{eq-exA}
(y_u, y_c, y_d, y_s, y_b)/y_t \sim&\ (\eps^3/t, \eps, \eps^2/t^2, \eps^2, t)
\sim (5\times 10^{-5}, 0.06, 2\times 10^{-4}, 0.004, 5), 
\notag \\
(s_{12}, s_{23}, s_{13}) \sim&\ (1/t, \eps^2, \eps^2/t)
\sim (0.2, 0.0036, 0.0008),  
\end{align}
where $\eps \sim 0.06$ and $t \sim 4.5$.
Here, we define $s_{ij} \coloneqq \sin \theta_{ij}$, 
where $\theta_{ij}$ being the mixing angle between the $i$-th and $j$-th generation 
in the standard parametrization of the CKM matrix~\cite{Workman:2022ynf}. 
For quark masses, $y_d$ and $y_s$ are well explained in this setup, 
because of the $t^2$ difference between them. 
The other ones, $y_u$, $y_c$ and $y_b$, are predicted to be larger 
than the experimental values by an order of magnitude. 
These are resolved by $\order{0.1}$ values of $\alpha_1$ and $\beta_{21}$ 
for $y_u$ and $y_b$, respectively. 
The charm Yukawa $y_c$ is realized by a tuning between $\alpha_2$ and $\alpha_3$. 
The phase of $\alpha_3$ should be small (but non-zero) to keep the charm mass light, 
because it can not be canceled by real $\alpha_2$. 
In other words, the phase of $\alpha_2$ and $\alpha_3$ should be approximately aligned 
to explain the charm mass. 
Regarding the CKM angles, $s_{12}$ and $s_{13}$ are well explained 
by the pattern Eq.~\eqref{eq-exA}, 
while $s_{23} \sim 0.004$ is an order of magnitude smaller than the experimental value. 
This gap is explained by the numerical factor of $\order{\eps^2}$ in the modular form $Y^{(8)}_2 \sim (1, 10/\sqrt{3}\,\eps^2)$.

Altogether, the CKM angles and $y_d$, $y_s$ are well explained in this model, 
while there should be $\order{0.1}$ hierarchies 
in $\alpha_1/\alpha^2_3$ and $\beta_{21}/\alpha_3^2$ for $y_u/y_t$ and $y_b/y_t$ respectively, 
and the $\order{0.1}$ tuning between $\alpha_2$ and $\alpha_3$ for $y_c/y_t$. 
Although the small hierarchies of $\order{0.1}$ may be simply accidental, 
we will discuss the possible origin for the small hierarchy in Sec.~\ref{sec-S3}.

\subsection{Small $\tan\beta$: $p=1$}
\label{sec-mdlB}

Now we consider the case with $p=1$ and the bottom quark mass is suppressed by $\eps$. 
We assign the modular weights as 
\begin{align}
k_{q_1} = 4, \quad k_{q_2} = 4, \quad k_u = 2, 
\quad 
k_{d_1} = 5, \quad k_{d_2} = 3, \quad k_{d_3} = 1,  
\end{align}
so that the superpotential is given by
\begin{align}
W =&\ H_u \left\{ \alpha_1 q_1 \left(Y_3^{(6)} u^c\right)_1 
         + \alpha_2 \left(q_2 Y^{(6)}_3 u^c \right)_1 
            + \sum_{i_Y=1}^2 \alpha_3^{i_Y} \left(q_2 Y^{i_Y(6)}_{3^\pr} u^c \right)_1 
            \right\}  \\ \notag 
        & \quad 
          + H_d  
 \left\{ \beta_{11} q_1 \left(Y^{(9)}_{\hat{1}} d^c_1 \right)_1  
        + \sum_{i=1}^3 \beta_{2i} \left( q_2 Y^{(11-2i)}_{\hat{2}}d^c_i\right)_{1} 
\right\}.
\end{align}
The Yukawa matrices are given by 
\begin{align}
 Y_u = &\ 
\begin{pmatrix}
 \alpha_1 [Y^{(6)}_3]_1 & 
 \alpha_1 [Y^{(6)}_3]_3 & 
 \alpha_1 [Y^{(6)}_3]_2 \\ 
-2 \alpha_2 [Y^{(6)}_3]_1 & 
   \alpha_2 [Y^{(6)}_3]_3 +\sqrt{3}\alpha_3^{i_Y} [Y^{i_Y(6)}_{3^\pr}]_2 & 
   \alpha_2 [Y^{(6)}_3]_2 +\sqrt{3}\alpha_3^{i_Y} [Y^{i_Y(6)}_{3^\pr}]_3 \\ 
-2 \alpha_3^{i_Y} [Y^{i_Y(6)}_{3^\pr}]_1 &  
   \alpha_3^{i_Y} [Y^{i_Y(6)}_{3^\pr}]_3 -\sqrt{3}\alpha_2 [Y^{(6)}_{3}]_2 & 
   \alpha_3^{i_Y} [Y^{i_Y(6)}_{3^\pr}]_2 -\sqrt{3}\alpha_2 [Y^{(6)}_{3}]_3 \\ 
\end{pmatrix}, 
\\ \notag 
Y_d =&\
\begin{pmatrix}
\beta_{11} Y^{(9)}_{\hsg}     & 0 & 0 \\ 
\beta_{21} [Y^{(9)}_{\hdb}]_1 & 
\beta_{22} [Y^{(7)}_{\hdb}]_1 & \beta_{23}[Y^{(5)}_{\hdb}]_1  \\ 
\beta_{21} [Y^{(9)}_{\hdb}]_2 & 
\beta_{22} [Y^{(7)}_{\hdb}]_2 & \beta_{23}[Y^{(5)}_{\hdb}]_2  
\end{pmatrix}. 
\end{align}
In $Y_d$, $(1,2)$ and $(1,3)$ elements vanish because
there is no $\hat{1}$ representation for $k<9$. 
There are 8 coefficients in this setup. 
We assign the odd weights for the down-quarks, 
and hence there are the hatted-representations in $Y_d$. 
Thus we should consider the homogeneous group $\Gamma_4^\pr$ unlike the first model. 
Since this hierarchy is, in principle, compatible with $\tan\beta =1$, 
it is not mandatory to have the second Higgs doublet to explain the quark hierarchies. 
This indicates that this model could be directly applied to the SM without supersymmetry.

At the benchmark point, the parameters are given by 
$\tan\beta = 1.6358$, $\tau = 1.4944+2.6779i$,  $|\alpha_3| = 1.2683\times 10^{-3}$  
and 
\begin{align}
\frac{1}{|\alpha_3|} 
\begin{pmatrix} 
\alpha_{1} \\ \alpha_{2} \\ \alpha_{3}^1 \\ \alpha_3^2
\end{pmatrix} 
=&\  
\begin{pmatrix} 
-0.2674 \\ 1.7408 \\ e^{-3.1281i} \\ -1.4009 
\end{pmatrix} 
, \quad 
\frac{1}{|\alpha_3|} 
\begin{pmatrix} 
\beta_{11} \\ \beta_{21} \\ \beta_{22} \\ \beta_{23} 
\end{pmatrix} 
= 
\begin{pmatrix} 
-6.9026 \\ -0.1294 \\ 0.2800 \\ 0.4095 
\end{pmatrix}.  
\end{align}
The ratio of the largest and smallest absolute values of the coefficients is about $53$. 
The slightly larger ratio is necessary 
because the parameter $\eps\sim 0.03$ is smaller than the first case.

The hierarchical structure after the canonical normalization gives the masses and CKM angles as 
\begin{align}
\label{eq-exB}
(y_u, y_c, y_d, y_s, y_b)/y_t \sim&\ (\eps^3, \eps, \eps^3/t^{1/2}, \eps^3 t^{1/2}, \eps\, t^{3/2})
\sim (3\times 10^{-5}, 0.03, 1\times 10^{-5}, 
6\times 10^{-5}, 0.4), 
 \notag \\ 
(s_{12}, s_{23}, s_{13}) \sim&\ (1, \eps^2, \eps^2)
\sim (1, 9\times 10^{-4}, 9\times 10^{-4}), 
\end{align}
where $\eps \sim 0.03$ and $t \sim 5.4$. 
The hierarchy well explains most of the hierarchical patterns
except $y_c$, $y_b$ and $s_{23}$.  
Similarly to the first model, $y_c$ is suppressed by the cancellation between $\alpha_2$ and 
$\alpha_3$, and $y_b$ is suppressed by $\beta_{21}$. 
The CKM angle $s_{23}$ is enhanced by the ratio of the coefficients  
$\abs{\beta_{11}/\beta_{21}} \sim 50$. 
Note that the first (second) row in the Yukawa matrices $Y_u$ and $Y_d$ 
is predominantly the second (first) generation, 
as opposed to the first model.

\subsection{$S_3$ origins of the small hierarchies and spontaneous CP violation}
%\label{sec-S3}
\textbf{}\label{sec-S3}

The small hierarchical structure of the coefficients in previous two models
can be explained by another modular symmetry $S_3$ at $N=2$. 
The $S_3$ modular symmetry would be realized 
in models with extra dimensions, e.g. $T_2\times T_2$, 
where the first $T_2$ leads $S_4^\pr$ and the second one leads $S_3$.
In both models of the previous sections, we found the small hierarchies in the coefficients, 
\begin{align}
\label{eq-S3hie}
\abs{\alpha_1} \ll \abs{\alpha_2}, \abs{\alpha_3}, 
\quad 
 \abs{\beta_{11}} \gg \abs{\beta_{21}}, \abs{\beta_{22}}, \abs{\beta_{23}}.   
\end{align}
Here we omit the upper index $i_Y$ for the multiple modular forms~\footnote{
$\beta_{21}^2$ with $\order{1}$ will not be a problem 
because $Y^{2(8)}_2$ does not have $\order{1}$ element, 
and thus only gives minor impacts on the result. 
}. 
Similarly to $\Gamma_4^\pr \simeq S_4^\pr$, 
we can consider $\Gamma_2 \simeq S_3$ to explain this hierarchy, 
by another small parameter $\eps_2 \sim 0.1$ controlled by another modulus $\tau_2$. 
The irreducible representations have the following hierarchies at $\Ita_2 \gg 1$, 
\begin{align}
    Y_1 \sim 1, \quad Y_{1^\pr} \sim \eps_2, 
    \quad 
    Y_2 \sim 
    \begin{pmatrix}
        1 \\ \eps_2
    \end{pmatrix},  
\end{align}
in the basis with real $S$ and diagonal $T$. 
The explicit forms of the modular forms are shown in App.~\ref{app-S3}.

Now, we assign $d_i^c$, $q_1$ to the trivial singlet $1$ and 
$u^c$, $q_2$ to the non-trivial singlet $1^\prime$ under the $S_3$ symmetry. 
Then the Yukawa matrices are given by
\begin{align}
\label{eq-Y3hie}
    Y_u \propto  
    \begin{pmatrix}
        \eps_2 & \eps_2 & \eps_2 \\ 
        1 & 1 & 1 \\ 
         1 & 1 & 1
    \end{pmatrix}, 
\quad 
Y_d \propto
 \begin{pmatrix}
   1 & 1 & 1 \\
      \eps_2 & \eps_2 & \eps_2 \\ 
          \eps_2 & \eps_2 & \eps_2 \\ 
    \end{pmatrix}. 
\end{align}
This can explain the hierarchical pattern in Eq.~\eqref{eq-S3hie}.
For example, we can construct $Y_1$ and $Y_{1^\pr}$ by the modular forms of weight 6
as explicitly shown in App.~\ref{app-S3}. 
The hierarchical structure of the Yukawa couplings is essentially realized 
by $\mathbbm{Z}_4^T\times \mathbbm{Z}_2^{T^\pr}$ symmetry, 
where the second one is from the $S_3$ modular symmetry. 
The residual symmetry $\mathbb{Z}_6^T \subset \Gamma_6$ plays 
a similar role in the model of Ref.~\cite{Kikuchi:2023cap}.

The spontaneous CP violation may be induced 
from the modulus VEV of the $S_3$ symmetry at $\order{\eps_2^2}$. 
If all of the $\order{1}$ coefficients are real, 
the CKM phase is vanishing up to $\order{\eps^4}$ and $\order{\eps_2^2}$. 
After rotating the phases of the quarks as 
\begin{align}
  Q^T \to 
  \begin{pmatrix}
    e^{-2i\phi} Q_1\\ e^{-i(2\phi+\phi_2)} Q_2 \\ e^{-i\phi_2} Q_3  
  \end{pmatrix}, 
  \quad 
  u^c \to 
  \begin{pmatrix}
     e^{i\phi_2} u^c_1\\ e^{i(\phi+\phi_2)} u^c_2 \\  e^{-i(\phi-\phi_2)} u^c_3  
  \end{pmatrix}, 
  \quad 
  d^c \to d^c,  
\end{align}
where $\phi := \mathrm{Arg}(\eps)$ and $\phi_2 := \mathrm{Arg}(\eps_2)$, 
the phases of the Yukawa matrices are given by 
\begin{align}
 \mathrm{Arg}\left(Y_u\right) 
 =&\ 
 \begin{pmatrix}
2\phi_2 & 2\phi_2 & 2\phi_2 \\
0 & 0 & 0 \\ 
0 & 4\phi & 0 
 \end{pmatrix}, 
 \quad 
 \mathrm{Arg}\left(Y_d\right) = 0. 
\end{align}
The phase $2\phi_2$ does not contribute to the CKM phase as it is common in the first row, 
and the phase $4\phi$ contributes only to the diagonalization of $u^c$ up to $\order{\eps^4}$. 
Thus there is no CP violation by non-zero VEV of the modulus at the leading order. 
There are, however, the CP phases from $\mathrm{Re}(\tau_2)$ at $\order{\eps_2^2} \sim \order{0.01}$ 
uncontrolled by the $\mathbbm{Z}^T_2$ symmetry. 
This will naturally explain the $\order{0.01}$ misalignment 
of the phase of the $\order{1}$ coefficients 
in the models in the previous sections. 
Note that the similar effects at $\order{\eps^4}$ will be too small 
to explain the observed CKM phase.

\section{Summary} 
\label{sec-summary}

In this paper, we demonstrated that the hierarchical structure 
of the quark masses and the CKM matrix is realized in the modular flavor symmetry 
at the level $N=4$. 
The FN-like mechanism is realized due to the $\mathbbm{Z}_4^T$ symmetry 
with a small parameter $\eps$ where the modulus is assumed to be stabilized at $\Ita \gg 1$. 
We found that there are only four cases shown in Eq.~\eqref{eq-creps}
in which the observed hierarchical structure in Eq.~\eqref{eq-texture} is realized.

We then studied the two examples with different assignments of the modular weights for the quarks.
In both models, the quark hierarchical structures are realized with $\order{1}$ coefficients, 
although the small hierarchy shown in Eq.~\eqref{eq-S3hie} of $\order{0.1}$ is needed, 
as well as the cancellation between the parameters $\alpha_2$ and $\alpha_3$ 
to explain the charm mass. 
We proposed a way to understand the small hierarchical structure Eq.~\eqref{eq-S3hie} 
and the origin of the CKM phase 
by the existence of another modular flavor symmetry $S_3$. 
In this case, the hierarchical strucutre of the Yukawa couplings
are essentially realized by $\mathbbm{Z}_4^T\times \mathbbm{Z}_2^{T^\pr}$, 
where the second symmetry comes from the second modular symmetry $S_3$. 
It was also turned out that the factors from canonical normalization $2\,\Ita \sim 5$ 
play an important role, because of the assignments of the modular weights. 
For instance, the Cabbibo angle is explained by this factor in the first model.

\section*{Acknowledgment}
The authors would like to thank M.Tanimoto for useful comments. 
The work of J.K. is supported in part by the Institute for Basic Science (IBS-R018-D1). 
This work is supported in part by he Grant-in-Aid for Scientific Research from the
Ministry of Education, Science, Sports and Culture (MEXT), Japan 
No.\ 22K03601~(T.H.) and 18K13534~(J.K.). 
The work of Y.A. is supported by JSPS Overseas Research Fellowships.

\afterpage{\clearpage} 
\appendix 
\section{Details of $S_4^\pr$ and $S_3$ symmetries} 
\label{app-S4p}

\subsection{Modular forms in $S_4^\pr$} 
\label{app-modform}

We choose the basis in which $T$ is diagonal and $S$ is real. 
For the doublet and the triplet, the representation matrices are respectively 
given by 
\begin{align}
\label{eq-rhoSTr2}
 \rho_S(2) = \frac{1}{2}
\begin{pmatrix}
 -1 & \sqrt{3} \\ \sqrt{3} & 1 
\end{pmatrix}, 
\quad 
\rho_T(2) = 
\begin{pmatrix}
 1 & 0 \\ 0 & -1 
\end{pmatrix}, 
\end{align}
and 
\begin{align}
\label{eq-rhoSTr3}
 \rho_S(3) = -\frac{1}{2}
\begin{pmatrix}
 0 & \sqrt{2} & \sqrt{2} \\ 
\sqrt{2} & -1 & 1 \\ 
\sqrt{2} & 1 & -1 \\ 
\end{pmatrix}, 
\quad 
\rho_T(3)= 
\begin{pmatrix}
 -1 & 0 & 0 \\ 0 & -i & 0 \\ 0 & 0 & i 
\end{pmatrix}. 
\end{align}
The primed and hatted representations are related as 
\begin{align}
 \rho_S(r) =&\ -\rho_S(r^\pr) = -i \rho_S(\hat{r}) =  i \rho_S(\hat{r}^\pr), \\
 \rho_T(r) =&\ -\rho_T(r^\pr) =  i \rho_T(\hat{r}) = -i \rho_T(\hat{r}^\pr),  \\
\mathbb{I} = \rho_R(r) =&\  \rho_R(r^\pr) =  - \rho_R(\hat{r}) = - \rho_R(\hat{r}^\pr).
\end{align}

 The products of the non-trivial singlets are given by 
\begin{align}
\label{eq-prodsgl}
 1^\pr \otimes 1^\pr = \hsg \otimes \hsg^\pr = 1, 
\quad 
\hsg\otimes\hsg = \hsg^\pr \otimes \hsg^\pr = 1^\pr, 
\quad 
1^\pr \otimes \hsg^\pr = \hsg,     
\quad 
1^\pr \otimes \hsg = \hsg^\pr.  
\end{align}
The similar relations for the prime and hat are applied for the other representations. 
For the doublet, however, $2^\pr$ ($\hdb^\pr$) should be understood 
as $(u_2, -u_1)$ to be $2$ ($\hdb$). For instance, 
the products of a singlet $\chi$ and doublet $u = (u_1,u_2)$ are given by 
\begin{align}
\label{eq-2prime}
 1^\pr \otimes 2 = \hsg \otimes \hdb = \hsg^\pr \otimes 2 = 1^\pr \otimes \hdb =&\  \chi
\begin{pmatrix}
  u_2 \\ - u_1 
\end{pmatrix},  
\quad %\\ 
\hsg\otimes 2 =
\hsg^\pr \otimes \hdb =   
\chi
\begin{pmatrix}
u_1 \\ u_2
\end{pmatrix}.   
\end{align}
The products of $2$ and $3$ are given by 
\begin{align}
2(u)\otimes 2(v) =&\  
(u_1v_1+u_2v_2)_1\oplus (u_1v_2-u_2v_1)_{1^\pr} \oplus 
\begin{pmatrix}
 u_2 v_2-u_1v_1 \\ u_1v_2 + u_2v_1 
\end{pmatrix}_2, \\ \notag   
2(u)\otimes 3(\phi) =&\  
\begin{pmatrix}
-2u_1 \phi_1  \\ 
u_1\phi_2 - \sqrt{3} u_2 \phi_3 \\ 
u_1\phi_3 - \sqrt{3} u_2 \phi_2 \\ 
\end{pmatrix}_3
\oplus
\begin{pmatrix}
-2u_2 \phi_1 \\ 
u_2\phi_2 + \sqrt{3} u_1 \phi_3 \\ 
u_2\phi_3 + \sqrt{3} u_1 \phi_2 \\ 
\end{pmatrix}_{3^\pr}, 
\\ \notag 
3(\phi)\otimes 3(\psi) =&\ 
\left(\phi_1\psi_1+\phi_2\psi_3 +\phi_3\psi_2 \right)_1 
\oplus 
\begin{pmatrix}
 2\phi_1\psi_1-\phi_2\psi_3-\phi_3\psi_2 \\ 
 \sqrt{3}\left(\phi_2\psi_2+\phi_3\psi_3\right)
\end{pmatrix}_2 
\\ \notag 
&\quad \oplus
\begin{pmatrix}
\phi_2\psi_2-\phi_3\psi_3 \\ 
-\phi_3\psi_1-\phi_1\psi_3 \\   
 \phi_1\psi_2+\phi_2\psi_1 
\end{pmatrix}_3 
\oplus
\begin{pmatrix}
\phi_2\psi_3-\phi_3\psi_2 \\ 
\phi_1\psi_2-\phi_2\psi_1 \\
\phi_3\psi_1-\phi_1\psi_3
\end{pmatrix}_{3^\pr}.  
\end{align}
Those for the representations with prime and/or hat are 
formally the same but with prime and/or hat accordingly to Eq.~\eqref{eq-prodsgl}.  
Note that doublets with prime should be understood 
in the same way as in Eq.~\eqref{eq-2prime}.

\begin{table}[tb]
 \center
\caption{\label{tab-wgt}
The number of representations of the modular forms at the weight $k \le 11$ 
in the $S_4^\pr$ modular symmetry.   
The representations for odd weights should be understood as the hatted ones. 
}
\vspace{0.5cm}
\begin{tabular}[t]{c|ccccccccccc} \hline 
 weight &  1 & 2 & 3 & 4 & 5 & 6 & 7 & 8 & 9 & 10 & 11 \\ \hline\hline
 $1$     & 0 & 0 & 0 & 1 & 0 & 1 & 0 & 1 & 1 & 1 & 0   \\
 $1^\pr$ & 0 & 0 & 1 & 0 & 0 & 1 & 1 & 0 & 1 & 1 & 1    \\ 
 $2$     & 0 & 1 & 0 & 1 & 1 & 1 & 1 & 2 & 1 & 2 & 2   \\ 
 $3$     & 1 & 0 & 1 & 1 & 2 & 1 & 2 & 2 & 3 & 2 & 3   \\ 
 $3^\pr$ & 0 & 1 & 1 & 1 & 1 & 2 & 2 & 2 & 2 & 3 & 3   \\ 
\hline 
\end{tabular}
\end{table}

The modular forms used in the model are given by
\begin{align}
 Y^{(4)}_{2} =&\ 
\begin{pmatrix}
\eps^8-10\eps^4\theta^4 + \theta^8 \\ 
 4\sqrt{3}\epth{2}{2} \left(\theta^4 + \eps^4\right)  
\end{pmatrix},  
\ &
 Y^{(4)}_{3} =&\ 
\eps\theta\left(\eps^4-\theta^4\right)  
\begin{pmatrix}
 -\sqrt{2}\eps\theta \\  -\eps^2 \\ \theta^2 
\end{pmatrix},  
\\ \notag
Y^{(5)}_{\hdb}=&\  \eps\theta \left(\eps^4-\theta^4\right)
\begin{pmatrix}
2\sqrt{3}\epth{2}{2} \\ \eps^4 + \theta^4
\end{pmatrix}, 
\ & 
& 
\\ \notag 
Y^{(6)}_{1^\pr}  =&\ \epth{2}{2}\left(\eps^4-\theta^4\right)^2,  
\ &  
Y^{(6)}_{2}=&\  (\eps^8+14\epth{4}{4}+\theta^8)
\begin{pmatrix}
\eps^4+\theta^4 \\ 
-2\sqrt{3}\epth{2}{2} 
\end{pmatrix}, 
\\ \notag 
%\ & 
 Y^{(6)}_3 =&\ \eps\theta \left(\eps^4-\theta^4\right)
\begin{pmatrix}
-2\sqrt{2} \eps\theta(\eps^4+\theta^4) \\ 
\eps^2 \left(\eps^4-5\theta^4\right) \\
\theta^2 \left(5\eps^4-\theta^4\right)  
\end{pmatrix},  
& 
Y^{1(6)}_{3^\pr} =&\ \eps\theta \left(\eps^4-\theta^4\right)
\begin{pmatrix}
4\sqrt{2} \epth{3}{3} \\ 
-\theta^2 \left(3\eps^4 + \theta^4\right) \\
\eps^2 \left(\eps^4+3\theta^4\right)  
\end{pmatrix}, 
\\ \notag 
 Y^{2(6)}_{3^\pr} =&\
\begin{pmatrix}
(\eps^4-\theta^4)^3 \\ 
8\sqrt{2} \epth{5}{3}\left(\eps^4+3\theta^4\right)  \\ 
8\sqrt{2} \epth{3}{5}\left(3\eps^4+\theta^4\right)  \\ 
\end{pmatrix}, 
\ & & \\ \notag 
Y^{(7)}_{\hdb}=&\ -\eps\theta\left(\eps^4-\theta^4\right) 
\begin{pmatrix}
 -4\sqrt{3}\epth{2}{2}\left(\eps^4+\theta^4\right) \\ 
\eps^8-10\epth{4}{4}+\theta^8 
\end{pmatrix},     
\hspace{-1.5cm} & & \\ \notag 
 Y^{1(8)}_2 =&\   \frac{1}{\sqrt{3}} 
\begin{pmatrix}
\sqrt{3}\left(\eps^{16}-130\epth{8}{8}+\theta^{16}\right) \\ 
2\epth{2}{2}\left(5\eps^{12}+91\epth{8}{4}+91\epth{4}{8}+5\theta^{12}\right) 
\end{pmatrix}, 
\hspace{-2.5cm}& & 
\\ \notag 
 Y^{2(8)}_2 = &\  \epth{2}{2}\left(\eps^4-\theta^4\right)^2 
\begin{pmatrix}
2\sqrt{3} \epth{2}{2} \\ \eps^4+\theta^4 
\end{pmatrix}, 
& & 
\\ \notag 
 Y^{(9)}_{\hsg}=&\  \epth{3}{3}\left(\eps^4-\theta^4\right)^3, 
\quad &
 Y^{(9)}_{\hdb}=&\  \eps\theta
(\eps^4-\theta^4)(\eps^8+14\eps^4\theta^4+\theta^8)
\begin{pmatrix}
 2\sqrt{3}\epth{2}{2} \\ \eps^4 + \theta^4 
\end{pmatrix}, 
\end{align}
where the functions are normalized such that 
the absolute value of the largest element is unity. 
The number of representations at weight $k\le 11$ 
are listed in Table~\ref{tab-wgt} for reference. 
There are $2k+1$ independent modular functions at a weight $k$ at the level $N=4$.

\subsection{Modular forms in $S_3$}
\label{app-S3}

For completeness, we derive the modular forms of $S_3$ following Ref.~\cite{Kobayashi:2018vbk}.
We work on the basis of the representations given by  
\begin{align}
 \rho_S = \frac{1}{2}
\begin{pmatrix}
 -1 & -\sqrt{3} \\ -\sqrt{3}& 1
\end{pmatrix}, 
\quad 
\rho_T = 
\begin{pmatrix}
   1 & 0 \\ 0 & -1  
\end{pmatrix},
\end{align}
then the modular form at weight $2$ is given by 
\begin{align}
 Y_2^{(2)} = 
\begin{pmatrix}
 Y_1 \\ Y_2 
\end{pmatrix},  
\end{align}
where 
\begin{align}
 Y_1(\tau):=&\ \frac{i}{2\pi} \frac{d}{d\tau} 
\left[ \log\eta\left(\frac{\tau}{2}\right) +\log\eta\left(\frac{\tau+1}{2}\right) 
 - 2 \log \eta(2\tau)\right], \\
 Y_2(\tau):=&\ \frac{i\sqrt{3}}{2\pi} \frac{d}{d\tau} 
\left[ \log\eta\left(\frac{\tau}{2}\right) -\log\eta\left(\frac{\tau+1}{2}\right) \right]. 
\end{align}
Here, $\eta(\tau)$ is the Dedekind eta function. 
The $q$-expansions of these functions are given by 
\begin{align}
 Y_1(\tau)=&\ \frac{1}{8} + \sum_{n,m=1}^\infty q^{2nm} \left\{ 
  -2n + (2n-1)q^{-m} + 2n q^{-n} 
  \right\}
\sim  1/8 + 3q + 3q^2 + 12 q^3
,  \\ 
Y_2(\tau)= &\ \sqrt{3} q^{\frac{1}{2}} \sum_{n,m=1}^\infty (2n-1) q^{2nm-n-m}
\sim
 \sqrt{3} q^{1/2} \left( 1+4q + 6q^2 + 8q^3 + \cdots\right). 
\end{align}

The modular forms with higher weights are given by 
\begin{align}
 Y^{(4)}_1 = Y_1^2 + Y_2^2, 
\quad   
 Y^{(4)}_2 = 
\begin{pmatrix}
 Y_2^2 - Y_1^2 \\ 2Y_1 Y_2
\end{pmatrix}, 
\end{align}
for $k=4$, and 
\begin{align}
 Y^{(6)}_1 = Y_1^3 - 3Y_1 Y_2^2, 
\quad   
 Y^{(6)}_{1^\pr} = Y_2^3 - 3Y_1^2 Y_2,  
\quad 
 Y^{(6)}_{2} = (Y_1^2+Y_2^2)
\begin{pmatrix}
 Y_1 \\ Y_2 
\end{pmatrix},  
\end{align}
for $k=6$. 
As expected from the residual $\mathbbm{Z}_2^T$ symmetry, 
the modular forms have the hierarchical structure Eq.~\eqref{eq-S3hie} 
at $\tau \sim i\infty$ and $\abs{\eps_2} := \abs{q^{1/2}} \ll 1$. 
Thus the weight should be $k\ge 6$ to realize the hierarchy of the Yukawa matrices in Eq.~\eqref{eq-Y3hie}.

{\small
\bibliography{ref_modular} 
\bibliographystyle{JHEP} % bst file
}

\end{document}